\documentclass[twocolumn,aps,pra,amsmath,floatfix]{revtex4}
\usepackage{graphicx}
\begin{document}

\title{Doping-driven Mott transition in La$_{1-x}$Sr$_x$TiO$_3$ 
       via simultaneous electron and hole doping of $t_{2g}$ subbands }
\author{A. Liebsch } 
\affiliation{Institut f\"ur Festk\"orperforschung, Forschungszentrum
             J\"ulich, 52425 J\"ulich, Germany}
\date{\today}

\begin{abstract}
The insulator to metal transition in LaTiO$_3$ induced by La substitution via 
Sr is studied within multi-band exact diagonalization dynamical mean field 
theory at finite temperatures. It is shown that weak hole doping triggers a 
large interorbital charge transfer, with simultaneous electron and hole doping 
of $t_{2g}$ subbands. The transition is first-order and exhibits phase separation 
between insulator and metal. In the metallic phase, subband compressibilities 
become very large and have opposite signs. Electron doping gives rise to an 
interorbital charge flow in the same direction as hole doping. These results 
can be understood in terms of a strong orbital depolarization.
\\ \mbox{\hskip1cm}\\
DOI: 10.1103  \hfill              PACS numbers: 71.10.Fd, 71.27.+a, 71.30.+h
\end{abstract}
\maketitle

The remarkable sensitivity of the electronic and magnetic properties of 
transition metal oxides to small changes of parameters such as temperature, 
pressure, or impurity concentration has attracted wide attention during 
recent years~\cite{imada}. A phenomenon of particular interest is the
filling-controlled metal insulator transition in the vicinity of integer 
occupancies of partially filled valence bands. For instance, LaTiO$_3$ is 
known to be a Mott insulator which becomes metallic if La is replaced by a 
few percent of Sr ions~\cite{kumagai,yoshida}. This kind of doping-driven 
insulator to metal transition has been discussed theoretically in terms of 
single-band models~%
\cite{furukawa,pruschke,fisher,kotliar,ono,camjayi,macridin,werner,garcia,eckstein} 
or multi-band models based on degenerate orbitals~%
\cite{zoelfl,nekrasov,oudovenko}.  
To our knowledge, only one calculation for non-identical subbands exists 
so far~\cite{craco}.

It has recently been shown that orbital degrees of freedom 
play a crucial role in the Mott transition of materials consisting 
of non-equivalent partially occupied subbands. For example, whereas cubic 
SrVO$_3$ ($3d^1$ configuration) is metallic, iso-electronic LaTiO$_3$ 
exhibits a  Mott transition because non-cubic structural distortions 
lift the degeneracy among $t_{2g}$ subbands~\cite{pavarini}. 
Local Coulomb interactions strongly enhance this orbital polarization, 
so that the electronic structure of the insulating phase consists of a 
single nearly half-filled band split into lower and upper Hubbard peaks, 
and two nearly empty $t_{2g}$ subbands. Similarly, the monoclinic 
symmetry of the Mott insulator V$_2$O$_3$ ($3d^2$) yields
a pair of nearly half-filled subbands (with lower and upper Hubbard peaks) 
and a single nearly empty $t_{2g}$ band~\cite{keller,poteryaev}. 
Also, the insulating phase of the layered perovskite Ca$_2$RuO$_4$ 
($4d^4$) consists of a fully occupied $t_{2g}$ band of predominant $d_{xy}$ 
character and two half-filled $d_{xz,yz}$-like bands~\cite{liebsch}. 
In view of the nearly complete orbital polarization of the Mott phase of 
these systems it is evidently necessary to go beyond single-band or 
degenerate multi-band models to describe the effect of electron or hole 
doping. 

In the present paper we investigate the destruction of the Mott phase of 
LaTiO$_3$ due to replacement of small amounts of La by Sr. To account 
for the orbital degrees of freedom during hole doping we make use of a 
recent multi-band extension of dynamical mean field theory based on exact 
diagonalization~\cite{perroni}. This approach has been shown 
to be highly useful for the study of strong correlations in several 
transition metal oxides~\cite{liebsch,perroni,naco}. Since it does not 
suffer from sign problems the full Hund exchange including spin-flip and 
pair-exchange contributions can be taken into account. Also, relatively 
large Coulomb energies and low temperatures can be reached.        

Naively, one might expect that all subbands are affected similarly by 
the doping process, perhaps weighted by their relative occupancies.
The result of this work is a completely different picture. 
As we show below, the addition of only a few percent of holes in the 
$t_{2g}$ bands triggers a substantially larger interorbital charge 
rearrangement, with a strong flow of electrons from the nearly half-filled 
subband towards the nearly empty ones. In other words, weak ``external'' hole 
doping of the $t_{2g}$ bands causes a much larger ``internal'' simultaneous 
electron and hole doping of different subbands. Thus, the density-driven 
insulator to metal transition in La$_{1-x}$Sr$_x$TiO$_3$ is accompanied by 
a significant reduction of orbital polarization. We also show that electron 
doping of the $t_{2g}$ bands, which could be achieved by substitution of La 
by Zr, leads to a strong charge flow in the same direction as hole doping, 
i.e., from the nearly half-filled $t_{2g}$ band to the nearly empty ones. 

As shown by Pavarini {\it et al.}~\cite{pavarini}, the octahedral distortions 
due to the orthorhombic lattice of LaTiO$_3$ ensure that the $t_{2g}$ 
density of states is not diagonal within the $xy,xz,yz$ basis. It is
nonetheless possible to construct from the $t_{2g}$ orbitals a new basis 
which is nearly diagonal~\cite{pavarini2}. We denote 
this basis here as $a_g$ and $e'_g$. The corresponding densities of states 
obtained from {\it ab initio} calculations within the local 
density approximation (LDA) are shown in Fig.~1. The centroid of the $a_g$ 
density lies about 200~meV below that of the two nearly degenerate 
$e'_g$ contributions. Accordingly, the subband occupancies (per spin band) 
in the absence of correlation effects are $n_{a}\approx 0.23$ and 
$n_{e}\approx 0.135$. This $t_{2g}$ crystal field splitting compares well with 
recent experimental estimates based on X-ray studies~\cite{cwik,haverkort}. 

To account for local Coulomb interactions among the Ti $3d$ electrons we use 
finite temperature dynamical mean field theory 
(DMFT)~\cite{dmft1,dmft2,dmft3,today} combined 
with exact diagonalization (ED)~\cite{ed}. It was recently shown~\cite{perroni} 
that this method can be generalized to multi-band materials by computing only 
those excited states of the impurity Hamiltonian that are within a narrow 
range above the ground state, where the Boltzmann factor provides a suitable 
convergence criterion. In the present case, each of the three $t_{2g}$
impurity orbitals couples to two bath levels, giving a cluster size $n_s=9$ 
with maximum Hamiltonian dimension 15.876. 
Exploiting the sparseness of the impurity Hamiltonian, a limited number of 
excited states can be computed very efficiently by using the Arnoldi 
algorithm~\cite{arnoldi}.
Since local Coulomb and exchange interactions couple the baths of the three 
$t_{2g}$ orbitals, the level spacing between excited states is very small so 
that finite size effects are much smaller than for a single impurity level 
coupled to only two bath levels. 

\begin{figure}[t]
  \begin{center}
  \includegraphics[width=5.4cm,height=8cm,angle=-90]{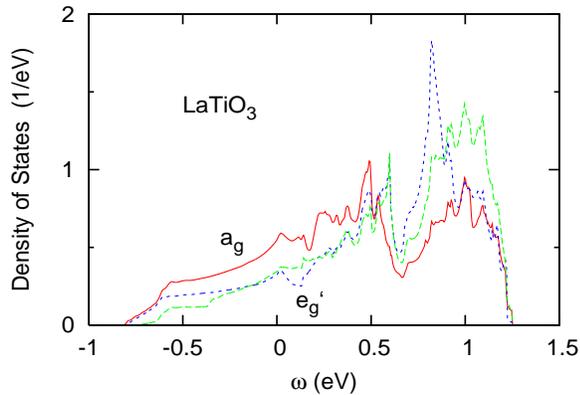}
  \end{center}
\vskip-4mm
\caption{(Color online) 
Density of states components of LaTiO$_3$ in nearly diagonal $t_{2g}$ subband 
representation~\cite{pavarini2}. Solid (red) curve: $a_g$ states, dashed 
(blue and green) curves: $e'_g$ states;  $E_F=0$.
}\end{figure}
         
Fig.~2 shows the variation of the $a_g$ and $e'_g$ subband occupancies with 
Coulomb energy. Full Hund exchange is included, and the exchange energy is 
held fixed at $J=0.65$~eV~\cite{pavarini}. The inter-orbital Coulomb energy 
is given by $U'=U-2J$. The temperature is $T=10\ldots20$~meV. For simplicity, 
the small difference between the $e'_g$ densities is neglected and their average 
density is used instead. Integer and non-integer occupancy are seen to lead
to qualitatively different behavior. For $n=2n_a+4n_e=1$, local Coulomb 
interactions gradually suppress orbital fluctuations up to $U\approx4.5$~eV. 
In the subsequent narrow range up to $U=5.0$~eV, the $a_g$ band is rapidly 
becoming nearly half-filled and the two $e'_g$ bands nearly empty. This result 
supports the picture presented in Refs.~\cite{pavarini,pavarini2}, where for 
$U=5.0$~eV, Ising exchange, and $T=100$~meV, LaTiO$_3$ was found to be a Mott 
insulator, with nearly empty $e'_g$ subbands, and the half-filled $a_g$ band 
split into lower and upper Hubbard peaks~\cite{craco2}. 

\begin{figure}[t]
  \begin{center}
  \includegraphics[width=5.4cm,height=8cm,angle=-90]{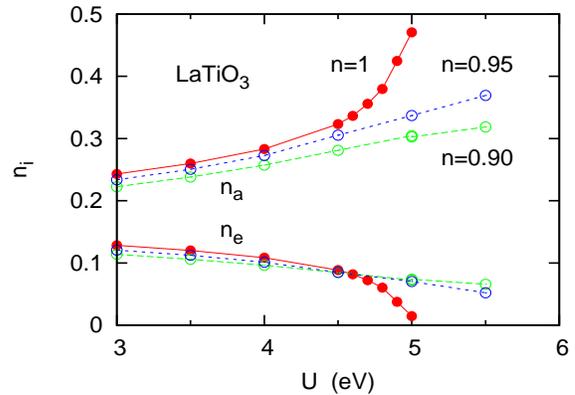}
  \end{center}
\vskip-4mm
\caption{(Color online) 
Subband occupancies (per spin band) of La$_{1-x}$Sr$_x$TiO$_3$ as  
functions of Coulomb energy. 
Solid (red) curves: $3d^1$ ($x=0$); dashed (blue) curves: $3d^{0.95}$ 
($x=0.05$); long-dashed (green) curves:  $3d^{0.90}$ ($x=0.1$).
}\end{figure}

For hole doping with $n=0.95$ and $0.90$, on the other hand, the orbital 
polarization increases smoothly with increasing Coulomb energy, without 
any sign of the critical behavior that is characteristic of the pure La 
compound. Evidently, the rapid suppression of orbital fluctuations near 
$U=5.0$~eV is a phenomenon associated with the integer occupancy of the 
$t_{2g}$ bands. To illustrate the destruction of the Mott phase of 
LaTiO$_3$ via doping with Sr we assume here that the primary effect is 
caused by the change of band filling. Additional modifications of the 
$a_g$ and $e'_g$ density of states due to the slightly smaller size of 
Sr compared to La are neglected. 

The unexpected and striking feature of these results is that the charge 
rearrangement among $t_{2g}$ orbitals is much larger than the number of  
holes induced via La $\rightarrow$ Sr substitution. For instance, for 
$x=0.05$ ($n=0.95$) the total $a_g$ occupancy (both spins) changes from 
near unity to $0.64$, whereas the total $e'_g$ occupancy (four spin bands) 
changes from near zero to $0.31$. Thus, the internal charge transfer from 
$e'_g$ to $a_g$ bands is six times larger than the external charge transfer 
due to doping with Sr. Clearly, the destruction of the Mott phase of LaTiO$_3$ 
via hole doping does not proceed via approximately equal participation of 
all subbands. Instead, weak external hole doping generates a much larger 
simultaneous electron and hole doping in different $t_{2g}$ subbands. 
           
Since the density of states of LaTiO$_3$ is not symmetric and the $t_{2g}$ 
occupation is far from one-half, it is interesting to consider also the 
case of electron doping, for example, via substituting La by Zr. 
Because the $a_g$ band is nearly full, one might expect the donated 
electrons to populate the nearly empty $e'_g$ bands, with little other 
changes. As in the hole doping case, we find an entirely different picture. 
For example, for $0.10$ electron doping, $0.26$ electrons are injected into 
the $e'_g$ bands and $0.16$ holes into the $a_g$ band. Thus, the actual 
number of electrons transfered to the $e'_g$ bands is $2.6$ times larger
than externally supplied. Of course, this can only be achieved via 
simultaneous hole doping of the nearly filled $a_g$ band.   

\begin{figure}[t]
  \begin{center}
  \includegraphics[width=5.4cm,height=8cm,angle=-90]{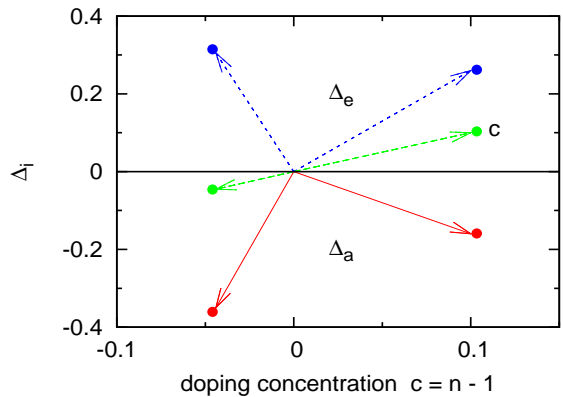}
  \end{center}
\vskip-4mm
\caption{(Color online) 
Red and blue dots: doping induced change of total $a_g$ and $e'_g$ subband 
occupancies, $\Delta_a$ and $\Delta_e$, respectively, for hole doping 
($c\approx-0.05$) and electron doping ($c\approx0.1$); $U=5.0$~eV.
In both cases, holes are injected into the nearly half-filled 
$a_g$ band (red arrows) and electrons into the nearly empty $e'_g$ bands 
(blue arrows). The green arrows denote the net doping.
}\end{figure}

These results are illustrated in Fig.~3 which shows the total electronic charge 
injected into the $e'_g$ bands ($\Delta_e = 4n_e>0$) and the total hole charge 
injected into the $a_g$ bands ($\Delta_a=2n_a-1<0$) for net doping concentrations 
$c\approx-0.05$ and $c\approx0.1$, where $c=\Delta_a+\Delta_e=n-1$. 
Both electron and hole doping are seen to cause a strong orbital depolarization, 
implying an internal charge flow between $t_{2g}$ bands that is significantly 
larger than the externally supplied number of electrons or holes. Note also the 
lack of symmetry between electron and hole doping.   
 
Fig.~4 shows in more detail the charge transfer among $t_{2g}$
bands as a function of chemical potential. The Coulomb and exchange energies
are held constant at values appropriate for LaTiO$_3$. Hole doping occurs for
$\mu<1.5$~eV, electron doping for $\mu>2.5$~eV. Because of the insulating gap,
the charge compressibility $\kappa=\partial n/\partial\mu$ vanishes in the 
intermediate range, i.e., the chemical potential can be varied without 
changing the subband occupancies. The small deviations from $n_a=0.5$ and 
$n_e=0.0$ in this region presumably are due to the finite temperature and ED 
finite size effects. The hysteresis loops near $\mu=1.5$~eV and $\mu=2.5$~eV 
demonstrate that the hole and electron driven Mott transitions at finite $T$
are first order. Phase separation, where an insulating solution with $n=1$ 
coexists with a metallic solution for $n\ne1$, is seen to occur at both transitions. 
Moreover, the compressibility gets very large as the upper and lower critical 
values of $\mu$ at both loops are approached. Note that the hysteresis behavior 
of the $a_g$ and $e'_g$ subbands is much more pronounced than that of the 
average $t_{2g}$ charge. Accordingly, the subband compressibilities 
$\kappa_i=\partial n_i/\partial \mu$ have opposite signs and their absolute 
values are much larger than the average $t_{2g}$ compressibility. 
These results show that the orbital degrees of freedom lead to a non-trivial 
generalization of the one-band picture close to half-filling~%
\cite{furukawa,pruschke,fisher,kotliar,ono,camjayi,macridin,werner,garcia,eckstein}
and of the multi-band picture based on equivalent orbitals~%
\cite{zoelfl,nekrasov,oudovenko}.

\begin{figure}[t]
  \begin{center}
  \includegraphics[width=5.4cm,height=8cm,angle=-90]{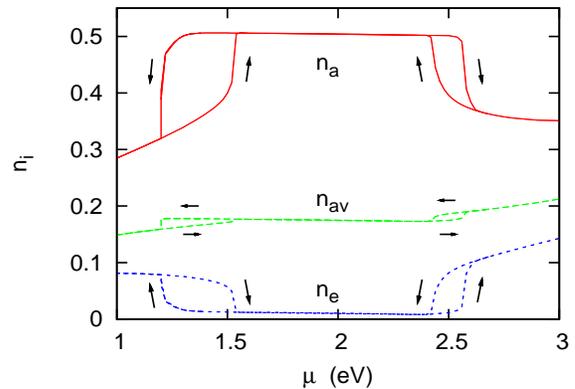}
  \end{center}
\vskip-4mm
\caption{(Color online) 
Solid (red) and dashed (blue) curves: $a_g$ and $e'_g$ subband occupancies 
(per spin band) of LaTiO$_3$ as functions of chemical potential for  
$U=5.0$~eV. Long-dashed (green) curve: average occupancy 
$n_{av}=(n_a+2n_e)/3$.
$\mu<1.5$~eV corresponds to hole doping, $\mu>2.5$~eV to electron doping.
The arrows mark the hysteresis behavior for increasing and decreasing $\mu$, 
indicating the first-order nature of the density-driven Mott transition.   
}\end{figure}

\begin{figure}[b]
  \begin{center}
  \includegraphics[width=4.0cm,height=8cm,angle=-90]{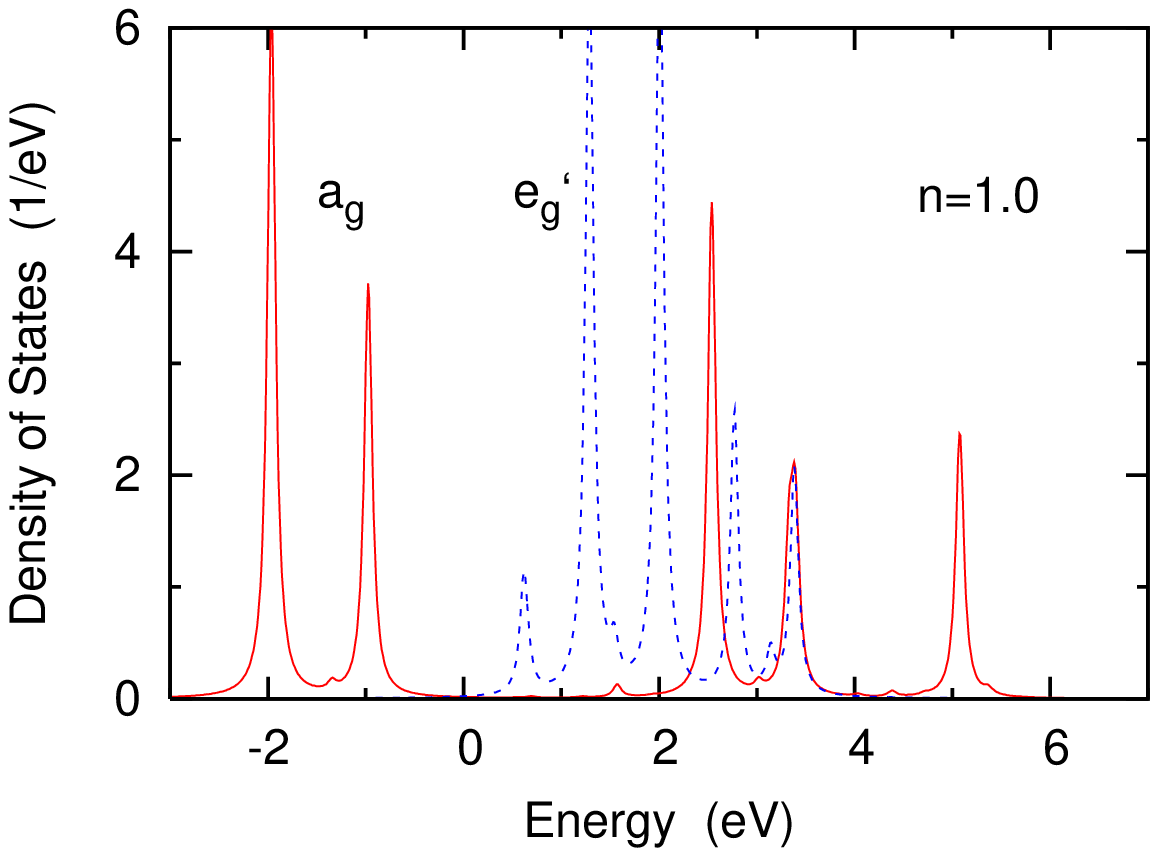}
  \includegraphics[width=4.0cm,height=8cm,angle=-90]{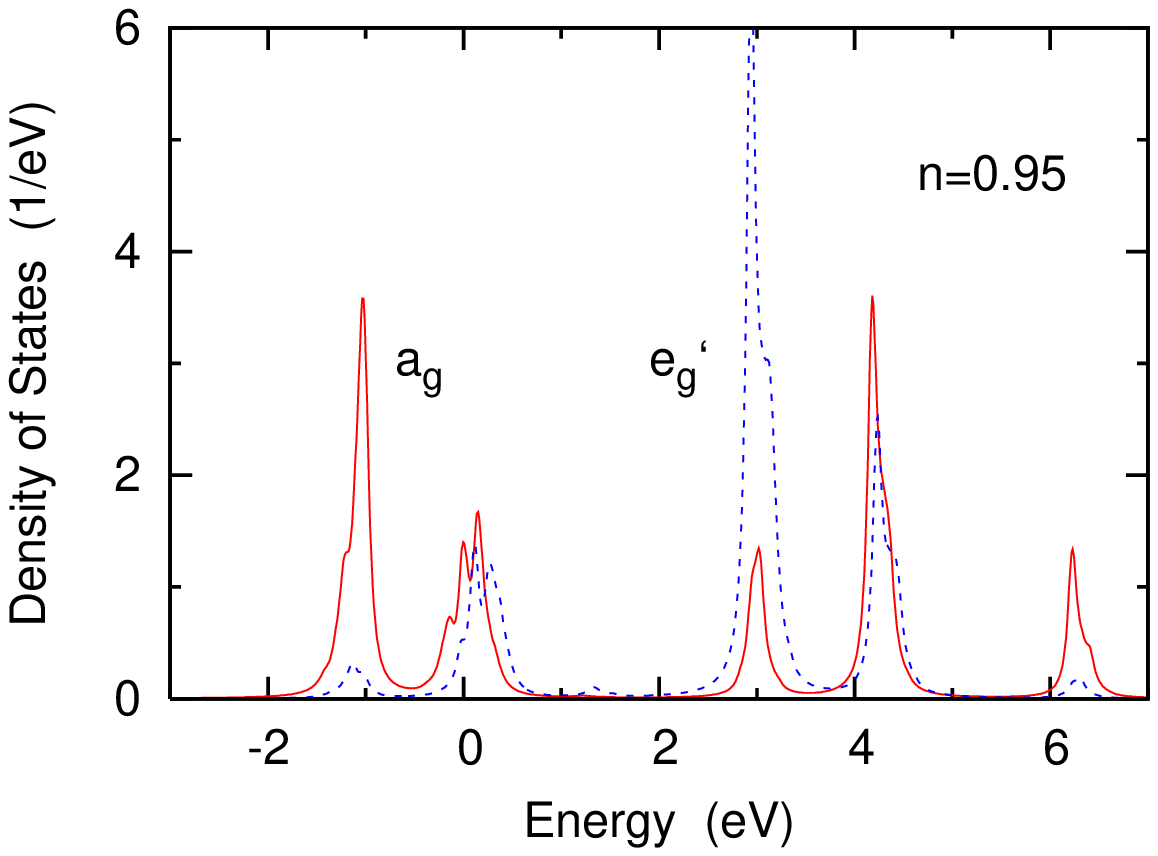}
  \end{center}
\vskip-4mm
\caption{(Color online) 
Quasi-particle spectra of LaTiO$_3$ (upper panel) and 
La$_{0.95}$Sr$_{0.05}$TiO$_3$ (lower panel) for $U=5$~eV. 
Solid (red) curves: $a_g$ states, dashed (blue) curves: $e'_g$ states;  $E_F=0$.
}\end{figure}

Finally, to make contact with the photoemission data for La$_{1-x}$Sr$_x$TiO$_3$%
~\cite{yoshida}, we show in Fig.~5 the quasi-particle spectra for $n=1$ ($x=0$) 
and $n=0.95$ ($x=0.05$). For simplicity, we give here the ED cluster spectra
which can be evaluated directly at real $\omega$, without requiring
extrapolation from Matsubara frequencies. 
At integer occupancy the system is insulating. The excitation gap is formed 
between the lower Hubbard peak of the $a_g$ band and the empty $e'_g$ bands, 
in agreement with Ref.~\cite{pavarini2}. The hole doped system, on the other 
hand, is metallic, with conduction states stemming from both $a_g$ 
and $e'_g$ bands. If we associate the peak near $E_F$ with 
coherent states, the large peak at $1.1$~eV binding energy, i.e., below the 
bottom of the LDA density of states, represents the incoherent weight related 
to the lower Hubbard band. It has mainly $a_g$ character, with weak $e'_g$ 
admixture. The position of this feature is in good agreement with the 
photoemission spectra for La$_{0.95}$Sr$_{0.05}$TiO$_3$~\cite{yoshida}. 
Moreover, the ratio between incoherent and coherent weight is roughly $3:1$. 
This is also consistent with the photoemission data which reveal substantially 
more incoherent than coherent emission. 

In summary, we have shown that multi-band ED DMFT is a powerful method to study 
the subtle interorbital charge rearrangement triggered by the density-driven 
Mott transition in La$_{1-x}$Sr$_{x}$TiO$_3$. In particular, we have shown that 
doping with Sr gives rise to a first-order insulator to metal transition that 
is accompanied by a significant orbital depolarization. Injection of only a few 
percent of holes into the Ti $t_{2g}$ bands leads to a much larger simultaneous 
electron and hole doping of different subbands. The transition exhibits 
separation between insulating and metallic phases. In the latter, subband 
charge compressibilities are very large and have opposite signs. Moreover, 
electron doping generates an internal charge flow in the same direction as 
hole doping. Since other Mott insulators such as V$_2$O$_3$ and 
Ca$_{2-x}$Sr$_{x}$RuO$_4$ 
also exhibit nearly complete orbital polarization, it is likely that doping 
these systems with electron or hole donors will lead to a similar enhancement 
of orbital fluctuations as discussed here for La$_{1-x}$Sr$_{x}$TiO$_3$.  

I like to thank Eva Pavarini for the density of states distributions shown
in Fig.~1, and Theo Costi for useful discussions.



\begin{thebibliography}{99}
\bibitem{craco2}
    In contrast, in Ref.~\cite{craco} orbital polarization was found to 
    slightly decrease in the insulating phase.
\end{thebibliography}
\end{document}